\documentclass[reprint, amsmath, amssymb, aps, physrev,nofootinbib]{revtex4-2}

\usepackage[T1]{fontenc}
\usepackage{csquotes}
\usepackage{graphicx,color} 
\usepackage{xcolor}
\usepackage{rotating}
\usepackage{amssymb}
\usepackage{amsfonts}
\usepackage{amsmath}
\usepackage{xspace}
\usepackage{mathtools} 
\usepackage{verbatim}
\usepackage{soul}
\usepackage[normalem]{ulem}
\usepackage{siunitx}
\usepackage{hyperref}
\hypersetup{hidelinks}
\usepackage{cleveref}
\crefname{equation}{Eq.}{Eqs.}
\crefname{figure}{Fig.}{Figs.}
\crefname{section}{Sec.}{Secs.}
\usepackage{breakcites}
\usepackage{accents}
\usepackage{dsfont}

\usepackage{listings}
\makeindex

\usepackage[thinlines]{easytable}

\usepackage[super]{nth}
\usepackage{braket}

\newcommand{\MP}{M_{\rm P}}
\newcommand{\Mass}{m}
\newcommand{\BareMass}{m_{\rm B}}
\newcommand{\BareCoupling}{\lambda_{\rm B}}
\newcommand{\PhysicalCoupling}{\lambda}
\newcommand\ie{\mbox{\textit{i.\,e.}}\xspace}
 
\newcommand\eg{\mbox{e.\,g.}\xspace}
\newcommand\D{\mathrm{d}}
\newcommand\Ord{\mathcal{O}}

\newcommand{\Prop}{D}
\newcommand{\FormFactor}{\mathcal{F}}
\newcommand{\landau}{\Lambda_{\rm LP}}
\newcommand{\NLScale}{\Lambda_{\rm NL}}

\begin{document}

\title{Finite is not enough: Triviality bounds in nonlocal field theories}




\author{Fabian Wagner}
\email{f.wagner@thphys.uni-heidelberg.de}
\affiliation{Institute for Theoretical Physics, Heidelberg University, Philosophenweg 16, 69120 Heidelberg, Germany}

\begin{abstract}
An ultraviolet-finite theory need not be ultraviolet complete. This observation weakens the case for spacetime discreteness in quantum gravity if that case rests on removing ultraviolet divergences alone. We arrive at this observation by deriving the one-loop momentum-subtraction flow in four-dimensional $\phi^4$ theories with ultraviolet-suppressing form factors. The suppression freezes the running above the nonlocality scale, but the theory can hit a Landau pole before suppression sets in. We obtain a bound on the infrared coupling by requiring perturbative control at all scales. This bound varies only at the percent level across the examples we consider. We expect similar bounds in models of discrete spacetime. Thus, requiring ultraviolet completeness improves the predictivity of nonlocal theories.
\end{abstract}


\maketitle

\section{Introduction}
\label{sec:intro} 
Quantum-gravity approaches broadly fall into two classes, namely those in which spacetime is fundamentally discrete and those in which spacetime is continuous.\footnote{Discreteness can also be a regulator in calculations, e.g., in (Causal) Dynamical Triangulations \cite{Ambjorn:1985az,Ambjorn:1998xu,Ambjorn:2026prt}.} A chief motivation for spacetime discreteness comes from ultraviolet (UV) divergences in quantum field theory (QFT). It is sometimes stated that a minimal length, implemented through spacetime discreteness or nonlocal propagation, removes UV divergences. Such a theory is UV finite; but is it also UV complete?

UV completeness is particularly important if quantum gravity is expected to resolve the Landau-pole problem of the Standard Model (SM). Extrapolating the SM hypercharge coupling gives a Landau-pole scale $\landau\sim \SI{1e42}{\giga\electronvolt}$ at three loops \cite{Buttazzo:2013uya}. This is far above the Planck scale, so its UV scaling requires accounting for quantum gravity. This means that a UV-complete theory of quantum gravity may also resolve the Landau pole problem of the SM. There is precedent: In asymptotic safety, there is evidence that gravity provides screening contributions to the running of potentially problematic couplings \cite{Harst:2011zx,Eichhorn:2017lry}. In string theory, the point-particle description responsible for the Landau pole gives way to string excitations at the string scale, whose interactions soften the ultraviolet behavior \cite{Polchinski:1998rq}. By contrast, a phenomenological Euclidean model of scalar propagation inspired by causal sets accelerates the running and exacerbates the Landau-pole problem \cite{deBrito:2023nie}.

In the present paper, we show that Landau poles can persist for a large class of nonlocal form factors that suppress the renormalization group (RG) flow above a nonlocality scale $\NLScale$. This highlights that spacetime discreteness as a fundamental principle may not be sufficient to obtain a UV-complete theory.  In addition, the infrared (IR) values of couplings must satisfy bounds to maintain perturbative control at all scales. These bounds resemble triviality bounds in effective field theory (EFT), which follow from requiring perturbative control up to a prescribed cutoff \cite{Luscher:1987ay}. Here, however, they constrain a putative fundamental theory. We call them \textit{weak triviality bounds} because they permit nonzero interactions at finite nonlocality scale but enforce perturbative triviality in the local limit $\NLScale\to\infty$.

We quantify the weak triviality bound for scalar $\phi^4$ theory in $d=4$. Intuitively, the weak triviality bound should depend only weakly on the details of the nonlocal modification because the coupling undergoes approximately local running over a wide range of scales before this running stops near $\NLScale$. Therefore, the nonlocal modification changes only a small part of the accumulated running. Indeed, we find that the bound's leading logarithm reproduces the EFT triviality bound with a cutoff at the nonlocality scale \cite{Luscher:1987ay}, which is independent of the form factor. In the examples we consider, the form-factor-dependent finite terms contribute to the bound only at the percent level for typical scale hierarchies, like Planckian nonlocality $\NLScale=\MP$ and evaluation at the electroweak scale $\mu=\SI{246}{\giga\electronvolt}$. In other words, the weak triviality bound is largely universal among nonlocal UV completions even at the quantitative level, and likely also applies to discrete-spacetime models. Thus, if spacetime discreteness is fundamental, EFT triviality bounds equally constrain the space of consistent fundamentally discrete theories.

This paper is structured as follows. In \cref{sec:review}, we review the triviality problem in scalar $\phi^4$ theory. \Cref{sec:Setup} defines the general class of nonlocal theories we study. In \cref{sec:Results}, we derive the MOM flow and weak triviality bound, establish its universal leading logarithm, and discuss the limitations of our approach. We conclude in \cref{sec:concl}. \Cref{app:examples} presents three example form factors, which we use in \cref{sec:Results}.

\section{Review of scalar $\phi^4$ theory and its triviality problem\label{sec:review}}

In a QFT, couplings \emph{run}, i.e., they depend on the momentum scale. Physically, quantum fluctuations make interactions momentum dependent. A central motivation for studying RG flows in perturbative QFT is that running couplings can capture this dependence for marginal interactions. In perturbative QFT, the RG flow can be derived in the process of regularizing UV divergences and renormalizing the theory. This might suggest the conclusion that UV divergences and the running of couplings are always intimately related. This need not be the case. Couplings run even in the absence of UV divergences in loop diagrams. As a consequence of the RG running, a different divergence can occur, namely one in the scale dependence of the coupling. Such a Landau pole can occur even if all loop diagrams are finite.

In UV-finite theories, the standard $\overline{\mathrm{MS}}$ scheme is misleading: without UV divergences, $\overline{\rm MS}$ couplings do not run by construction. Still, the one-particle-irreducible vertices remain momentum dependent. Similarly, the $\overline{\rm MS}$ scheme does not capture the full momentum dependence of higher-derivative theories such as quadratic gravity due to IR singularities \cite{Buccio:2024hys, Buccio:2024omv,Buccio:2025tci}.\footnote{In these theories, IR divergences contribute to the momentum dependence of amplitudes in addition to the UV divergences.} Other renormalization schemes such as the MOM scheme we use in this paper capture the vertex running more faithfully.

In local QFT, a Landau pole prohibits entire classes of interactions. Any coupling which feeds into the flow of a coupling that develops a Landau pole must vanish throughout the whole RG trajectory. In many cases, one ends up with a free theory. This is the triviality problem \cite{Landau:1954nau,Landau:1955ip,Pomeranchuk:1956zz,Landau:1956zr}. The same issue arises in the SM, affecting the Higgs self-interaction \cite{Lee:1977yc,Cabibbo:1979ay} and the Abelian hypercharge coupling \cite{Landau:1955ip}.

Local scalar $\phi^4$ theory in $d=4$ is trivial. The first hints of triviality came from perturbation theory \cite{Pomeranchuk:1956zz,Wilson:1973jj}: In local $\phi^4$ theory, the one-loop beta function for the quartic coupling $\PhysicalCoupling$ reads in the $\overline{\rm MS}$ scheme\footnote{Here, we normalize the coupling of the $\phi^4$ operator as $(\PhysicalCoupling/4!)$.}
\begin{equation}
	\beta_\PhysicalCoupling\equiv\mu\partial_\mu\PhysicalCoupling=\frac{3\PhysicalCoupling^2}{16\pi^2},\label{eq:LocalBeta}
\end{equation}
where $\mu$ is the scale at which the coupling is probed. This flow can be integrated analytically such that
\begin{equation}
	\PhysicalCoupling(\mu')=\frac{\PhysicalCoupling(\mu)}{1-\frac{3\PhysicalCoupling(\mu)}{16\pi^2}\log\frac{\mu'}{\mu}}.
\end{equation}
Thus, for every positive $\PhysicalCoupling(\mu)$ there is a UV scale $\mu'>\mu$ at which $\PhysicalCoupling(\mu')$ diverges. One can avoid this pole only if $\PhysicalCoupling(\mu)=0$ for all $\mu$. More precisely, within perturbation theory one can only show that perturbation theory must break down in the UV unless the theory is trivial at all scales. However, beyond perturbation theory, scalar $\phi^4$ theory in $d>4$ has been rigorously proven to be trivial \cite{Aizenman:1982ze}. This proof has recently been extended to $d=4$ \cite{Aizenman:2019yuo}, as previously suggested by nonperturbative lattice studies \cite{Frohlich:1982tw,Luscher:1987ay,Gockeler:1997dn}.

A trivial QFT requires a UV completion. While the theory can function as an EFT at scales below the Landau-pole scale $\landau$, it must be modified to avoid the triviality problem. For instance, the Landau pole in the SM can be resolved by extending the particle content, \eg, via vector-like fermions \cite{Litim:2014uca,Litim:2023tym} or supersymmetric grand unification \cite{Dimopoulos:1981yj,Giudice:2004tc,Hisano:2013cqa, Hall:2013eko}.

Nonlocal form factors can suppress the running of couplings above the nonlocality scale $\NLScale$ \cite{Evens:1990wf,Moffat:1990jj,Kleppe:1991rv,Ghoshal:2017egr,Frasca:2021iip}.  This suppression freezes the flow before a coupling can reach its Landau pole if $\landau\gg\NLScale$, which has motivated claims that nonlocal QFTs may avoid the Landau-pole problem altogether \cite{Green:2020rsb,Emmel:2021fsx}. In \cref{sec:Results}, we show that suppression alone is not sufficient to avoid a Landau pole. In addition, the strength of interactions must be bounded at finite scales. Before deriving this bound, we specify the class of nonlocal theories we work with in this paper.

\section{Nonlocal field theory\label{sec:Setup}}

Nonlocality can modify propagators, vertices, or both \cite{Efimov:1967pjn,Efimov:1969fd,Tomboulis:1997gg,Modesto:2011kw}; see \cite{Modesto:2017sdr,BasiBeneito:2022wux} for recent reviews. In this paper, we consider nonlocal modifications of the propagator only, while the interactions are local. The nonlocal modifications to the propagator are captured by the form factor $\FormFactor_p$, which depends on the Euclidean norm of the momentum $p$. We prescribe this form factor in the bare action, and do not derive it from quantum corrections. The resulting kinetic operator must not be a finite polynomial in derivatives for the theory to be truly nonlocal.

The form factor $\FormFactor_p$ monotonically interpolates between 1 when\footnote{Strictly speaking, for some form factors one also requires $\Mass/\NLScale\to0$ with mass $\Mass$, see \cref{eq:GeneralPropagator}.} $p/\NLScale\to0$ and 0 when $p/\NLScale\to\infty$, where it decays fast enough to render all amplitudes finite. We further require that $\FormFactor$ has no poles in the complex plane. This is important because additional poles often add ghostly degrees of freedom. Typical nonlocal form factors in the literature generally share these properties \cite{Padmanabhan:1996ap,Tomboulis:1997gg,Modesto:2011kw,Biswas:2011ar,Modesto:2014lga,Buoninfante:2018mre,Buoninfante:2018gce}.

Nonlocal field redefinitions can transfer nonlocal operators from the kinetic term to the interactions \cite{Buoninfante:2018mre,Buoninfante:2022krn,BasiBeneito:2022wux}. As the form factor $\FormFactor_p$ has no real zeros, the two descriptions agree at the exact level if sources and the path-integral measure are transformed accordingly \cite{Buoninfante:2022krn,Cohen:2024fak}. Thus, a model with a nonlocal propagator is equivalent to a theory with nonlocal interactions.

The models we consider all belong to one family of actions. We study a real scalar $\phi^4$ theory in four Euclidean dimensions with a local $\phi^4$ interaction, \ie,
\begin{equation}
	\begin{aligned}
		 S=&\frac{1}{2}\int_{p}\phi_{\mathrm{B},p}
		 \Prop_{\rm B}^{-1}(p)\phi_{\mathrm{B},-p}\\
		 &+\frac{\BareCoupling}{4!}\int_{p_1,p_2,p_3}
		 \phi_{\mathrm{B},p_1}\phi_{\mathrm{B},p_2}\phi_{\mathrm{B},p_3}\phi_{\mathrm{B},-p_1-p_2-p_3},
		 \label{eq:BareAction}
	\end{aligned}
\end{equation}
where $\int_p=\int_{\mathbb{R}^4}\D^4p/(2\pi)^4$, $\BareCoupling\geq0$ is the bare quartic coupling, and $\phi_{\rm B}$ is the bare field. Wave-function renormalization starts at two loops, so we identify \(\phi_{\mathrm B}\) with the renormalized field $\phi$ at one loop. The bare propagator reads
\begin{align}
 \Prop_{\rm B}(p)&=\frac{\FormFactor\left(p^2/\NLScale^2,
 (p^2+\BareMass^2)/\NLScale^2\right)}{p^2+\BareMass^2}.
 \label{eq:GeneralPropagator}
\end{align}
Here, $\FormFactor_{\rm B,p}\equiv\FormFactor(p^2/\NLScale^2,
(p^2+\BareMass^2)/\NLScale^2)$ is the bare form factor, and $\BareMass$ is the bare mass.

We provide three examples of widely used form factors and some of their motivation in \cref{app:examples}, and compare them in \cref{sec:FF_dep}. First, we keep $\FormFactor_p$  general.

\section{Weak triviality problem in finite theories}\label{sec:Results}\label{Sec:VanishingMass}

We now show that a UV-finite theory need not be free from the triviality problem. Finite loop integrals depend on external momenta. Therefore, they can drive a theory towards a strong-coupling regime. We start by calculating the finite one-loop four-point vertex for a general form factor $\FormFactor$.

\subsection{One-loop four-point vertex}

The one-loop momentum dependence of the four-point vertex is determined by three channels. For incoming momenta with $\sum_I p_I=0$, we define the Euclidean Mandelstam variables
\begin{align}
 s\equiv(p_1+p_2)^2,&&
 t\equiv(p_1+p_3)^2,&&
 u\equiv(p_1+p_4)^2.
\end{align}
Each channel contributes the finite bubble integral
\begin{equation}
 B(P^2)=\int_q
 \frac{\FormFactor_{{\rm B},q}\FormFactor_{{\rm B},q+P}}
 {(q^2+\BareMass^2)[(q+P)^2+\BareMass^2]}.
 \label{eq:bubble_integral_B}
\end{equation}
The vertex $\Gamma^{(4)}$ is then at one loop
\begin{equation}
	\begin{aligned}
		 \Gamma^{(4)}(s,t,u)=\BareCoupling
		 -&\frac{\BareCoupling^2}{2}
		 \left[B(s)+B(t)+B(u)\right]\\
		 & +\Ord(\BareCoupling^3).
	 \end{aligned}
 \label{eq:OneLoopVertex}
\end{equation}
At one loop, mass and field renormalization do not affect the four-point amplitude: the one-loop mass shift enters the bubble contribution only at $\Ord(\BareCoupling^3)$, while field renormalization starts at two loops. We can therefore replace the bare mass with the pole mass $m$ such that
\begin{align}
 	B(P^2)=\int_q
	\frac{\FormFactor_q\FormFactor_{q+P}}
	{(q^2+m^2)[(q+P)^2+m^2]}+\Ord(\BareCoupling).
	\label{eq:bubble_integral}
\end{align}

The $\overline{\rm MS}$ scheme cannot track the momentum dependence of \cref{eq:OneLoopVertex} because the bubble is not UV divergent. Therefore, we now define an operational MOM scheme.

\subsection{Momentum subtraction\label{sec:MOM}}

MOM schemes fix one Euclidean momentum configuration which is parametrized by one scale $\mu$. Different momentum configurations define different schemes. In \cref{sec:SchemeIndependence}, we show that the leading logarithm is scheme independent for $\max(m,\mu)\ll\NLScale$, where $\mu$ is the RG scale.

We choose the symmetric Euclidean point which treats all three channels in \cref{eq:OneLoopVertex} equally. For external momenta $p_I$, the symmetric Euclidean point is defined by
\begin{equation}
 p_I^2=\mu^2,\qquad
 p_I\cdot p_J=-\frac{\mu^2}{3}\quad(I\neq J),
 \label{eq:SymmetricPoint}
\end{equation}
where $I,J=1,2,3,4$. \Cref{eq:SymmetricPoint} implies
\begin{equation}
 s=t=u=\frac{4\mu^2}{3}.
\end{equation}
Thus, each channel carries the momentum $|P|=2\mu/\sqrt{3}$.

We define the MOM coupling $\PhysicalCoupling(\mu)$ by the vertex at the symmetric point. Its one-loop relation to the bare coupling is
\begin{align}
 \PhysicalCoupling(\mu)\equiv\,& \Gamma^{(4)}\left(\frac{4\mu^2}{3},\frac{4\mu^2}{3},\frac{4\mu^2}{3}\right),\\
 =\,&\BareCoupling
 -\frac{3\BareCoupling^2}{2}
 B\left(\frac{4\mu^2}{3}\right)
 +\Ord(\BareCoupling^3).
 \label{eq:MOMDefinition}
\end{align}
Inverting this relation perturbatively yields
\begin{equation}
 \BareCoupling=\PhysicalCoupling(\mu)
 +\frac{3\PhysicalCoupling^2(\mu)}{2}
 B\!\left(\frac{4\mu^2}{3}\right)
 +\Ord(\PhysicalCoupling^3).
 \label{eq:BareFromMOM}
\end{equation}
We can see already at this stage that if $\PhysicalCoupling(\mu)$ is of order $ 2/[3B(4\mu^2/3)]$, the one-loop corrections in \cref{eq:OneLoopVertex} are not suppressed relative to the tree-level amplitude, and perturbation theory breaks down.

In terms of the MOM coupling, the loop correction to the four-point amplitude in \cref{eq:OneLoopVertex} vanishes at the symmetric point. Inserting \cref{eq:BareFromMOM} into \cref{eq:OneLoopVertex} yields
\begin{equation}
	\begin{aligned}
		 \Gamma^{(4)}(s,t,u)={}\PhysicalCoupling(\mu)
		 &-\frac{\PhysicalCoupling^2(\mu)}{2}\bigg[
		 B(s)+B(t)+B(u)\\
		 &-3B\!\left(\frac{4\mu^2}{3}\right)\bigg]
		 +\Ord(\PhysicalCoupling^3).
		 \label{eq:SubtractedVertex}
	\end{aligned}
\end{equation}
Equation~\eqref{eq:SubtractedVertex} makes the subtraction scale explicit. The $\mu$-dependences of the coupling and the subtracted loop term cancel each other.

The bare coupling does not depend on $\mu$. We therefore differentiate \cref{eq:BareFromMOM} at fixed $\BareCoupling$, $m$, and $\NLScale$ to obtain
\begin{equation}
 \beta_\PhysicalCoupling(\mu)
 =-\frac{3\PhysicalCoupling^2(\mu)}{2}
\mu\partial_\mu
 B\!\left(\frac{4\mu^2}{3}\right)
 +\Ord(\PhysicalCoupling^3).
 \label{eq:MOMBeta}
\end{equation}

In the local limit $\NLScale\to\infty$ and for $\mu\gg m$, \cref{eq:MOMBeta} recovers \cref{eq:LocalBeta}. Let $B_0$ denote the bubble with $\FormFactor=1$ and an auxiliary UV regulator. Feynman parametrization yields, after differentiating and removing that regulator,
\begin{equation}
 \frac{\D B_0(P^2)}{\D\text{$\log |P|$}}
 =-\frac{1}{8\pi^2}\int_0^1\D x\,
 \frac{x(1-x)P^2}{m^2+x(1-x)P^2}.
 \label{eq:LocalBubbleDerivative}
\end{equation}
At the symmetric point, $|P|=2\mu/\sqrt3$ and $\D/\D\log\mu=\D/\D\log |P|$. For $\mu\gg m$, the integral approaches one and \cref{eq:MOMBeta} recovers \cref{eq:LocalBeta}. For $\mu\ll m$, the integral scales as $P^2/(6m^2)=2\mu^2/(9m^2)$ and implements the usual mass threshold.
We next integrate the flow in \cref{eq:MOMBeta} at finite $\NLScale$.

\subsection{Weak triviality\label{sec:WeakTriviality}}

We solve \cref{eq:MOMBeta} analytically, obtaining
\begin{equation}
\PhysicalCoupling(\mu)
 =\frac{\PhysicalCoupling(\mu')}{1+\frac{3\PhysicalCoupling(\mu')}{2}\left[
 	B\!\left(\frac{4\mu^2}{3}\right)
 	-B\!\left(\frac{4\mu'^2}{3}\right)
 	\right]}.
 \label{eq:IntegratedMOM}
\end{equation}
As long as perturbation theory applies, we extend the running in \cref{eq:IntegratedMOM} to arbitrarily high scales. We define the UV limit
\begin{equation}
	\PhysicalCoupling_{\rm UV}=\lim_{\mu\to\infty}\PhysicalCoupling(\mu).
\end{equation}
For the form factors considered here, $\lim_{\mu\to\infty}B(\mu^2)=0$. Therefore, evaluating the upper scale in \cref{eq:IntegratedMOM} at infinity yields
\begin{equation}
\PhysicalCoupling_{\rm UV}
 =\frac{\PhysicalCoupling(\mu)}{1-
 \frac{3\PhysicalCoupling(\mu)}{2}
 B\!\left(\frac{4\mu^2}{3}\right)}
.
 \label{eq:MOMMaster}
\end{equation}

The UV limit $\PhysicalCoupling_{\rm UV}$ bounds $\PhysicalCoupling(\mu)$ at all scales. For positive, monotonically decreasing form factors $\FormFactor_p$, $B(\mu^2)$ also decreases monotonically with $\mu^2$. Consequently, $\lambda(\mu)$ in \cref{eq:IntegratedMOM} is monotonically increasing and reaches its maximum as $\mu\to\infty$. 

Within perturbation theory, we cannot diagnose a Landau pole. However, we can identify RG trajectories which become strongly coupled, \ie, along which perturbation theory ultimately breaks down. We choose a finite maximum coupling $\PhysicalCoupling_{\rm crit}$, which indicates the breakdown of perturbation theory. As there is no universal sharp boundary for this breakdown, we leave the exact value of $\PhysicalCoupling_{\rm crit}$ unspecified; common choices are $\PhysicalCoupling_{\rm crit}=1$ and $\PhysicalCoupling_{\rm crit}=16\pi^2$. The theory can only be perturbatively UV complete if $\PhysicalCoupling_{\rm UV}\leq\PhysicalCoupling_{\rm crit}$. As the RG trajectory is monotonic, $\PhysicalCoupling(\mu)<\PhysicalCoupling_{\rm crit}$ for all $\mu $. Consequently, \cref{eq:MOMMaster} yields the perturbative bound
\begin{equation}
 \PhysicalCoupling_{\rm pert}(\mu;\PhysicalCoupling_{\rm crit})\equiv\frac{\PhysicalCoupling_{\rm crit}}{1+\frac{3\PhysicalCoupling_{\rm crit}}{2}B\left(\frac{4\mu^2}{3}\right)}.
 \label{eq:PerturbativeCriticalCoupling}
\end{equation}
The sign in the denominator has changed in comparison to \cref{eq:MOMMaster} because here we write an IR coupling in terms of a UV ceiling, while \cref{eq:MOMMaster} describes the UV coupling in terms of an IR boundary condition.

All boundary values $0\leq\PhysicalCoupling(\mu)\leq\PhysicalCoupling_{\rm pert}(\mu;\PhysicalCoupling_{\rm crit})$ remain perturbative throughout the RG flow. Boundary values $\PhysicalCoupling(\mu)>\PhysicalCoupling_{\rm pert}(\mu;\PhysicalCoupling_{\rm crit})$ lead to a strongly coupled theory at sufficiently large scales.

We call \cref{eq:PerturbativeCriticalCoupling} a weak triviality bound. At fixed nonlocality scale, the coupling of the fundamental finite theory need not vanish, but requiring perturbative control at all scales constrains its strength. We stress that the weak triviality bound is not an EFT bound, but constrains a fundamental theory.

In summary, to avoid a strong-coupling regime, we have to bound $\PhysicalCoupling(\mu)$ from above. This indicates that the finite theory suffers from a weaker version of the triviality problem, where interactions are allowed but their strength is constrained. 

\subsection{Leading logarithm\label{sec:UniversalLimit}}

We now show that the leading logarithmic contribution to $\PhysicalCoupling_{\rm pert}$ is universal across form factors $\FormFactor$ and nonexceptional MOM configurations. We consider masses and external momenta well below the physical suppression scale in $\FormFactor$, \ie, $\max(m,\mu)\ll\NLScale$. For a Planckian regulator, $\NLScale\sim\MP$, this hierarchy is generally satisfied in present-day experiments. We therefore now evaluate the leading logarithmic contribution in $\max(m,\mu)/\NLScale$.

The leading logarithmic contribution to \cref{eq:bubble_integral} stems from intermediate loop momenta. We divide the bubble integral up into three regions,
\begin{align}
	\mathcal{R}_{\rm IR}=&\{q,|q|<\max(m,|P|)\},\\
	\mathcal{R}_{\rm int}=&\{q,\max(m,|P|)<|q|<\NLScale\},\\
	\mathcal{R}_{\rm UV}=&\{q,\NLScale<|q|\}.
\end{align}
The corresponding integrals are
\begin{equation}
	\begin{aligned}
		&B_{\rm IR/int/UV}\equiv\\
		&\qquad\int_{\mathcal{R}_{\rm IR/int/UV}}\frac{\FormFactor_q\FormFactor_{q+P}}{(q^2+m^2)((q+P)^2+m^2)},\label{eq:def_B_IR_int_UV}
	\end{aligned}
\end{equation}
where the measure $\D^4q/(2\pi)^4$ is again implied. Then, $B=B_{\rm IR}+B_{\rm int}+B_{\rm UV}$. 

The integral $B_{\rm IR}$ is independent of $\NLScale$ at leading order. For $|q|\ll\NLScale$ the form factor is $\FormFactor_q\approx\FormFactor_{q+P}\approx1$. In this approximation, $B_{\rm IR}$ can be evaluated analytically and is finite because its integrand is dominated by the physical IR regulator $\max(m,|P|)$. Thus, $B_{\rm IR}$ depends on $\NLScale$  only at higher order in the expansion of $\FormFactor_q$, which is suppressed relative to the leading finite contribution. 

The integral $B_{\rm UV}$ is bounded for small $\max(m,|P|)/\NLScale$. In the UV, we can neglect the mass and the external momentum such that 
\begin{equation}
	B_{\rm UV}\approx\frac{1}{8\pi^2}\int_{1}^{\infty}\frac{\D z}{z}\FormFactor^2(z^2,z^2),
\end{equation}
where $z=|q|/\NLScale$. As $\FormFactor$ suppresses the UV, $B_{\rm UV}$ is finite, and, therefore, does not contribute a logarithm in $\max(m,\mu)/\NLScale$.

For $B_{\rm int}$, we can neglect $m/|q|$ and $|P|/|q|$ at leading order and approximate $\FormFactor_q\approx1$ for $|q|<\NLScale$. For this approximation, it is crucial that both $m,|P|<|q|$ in \cref{eq:def_B_IR_int_UV}, which is assured by the lower integration boundary of $B_{\rm int}$, namely $\max(m,|P|)$.  As a result, we obtain
\begin{align}
	B_{\rm int}(P^2)\approx&\frac{1}{8\pi^2}\int_{\max(m,|P|)}^{\NLScale}\frac{\D |q|}{|q|},\\
	=&\frac{1}{8\pi^2}\log\frac{\NLScale}{\max(m,|P|)}.
\end{align}

The remaining contributions to $B_{\rm int}$ are bounded as $\max(m,\mu)/\NLScale\to0$ under two assumptions. First, $\FormFactor(x,y)$ must approach $\FormFactor(0,0)=1$ at least as fast as a positive power of $x+y$. Second, $\FormFactor$ decreases monotonically. Both of these properties are satisfied by typical examples in the literature \cite{Padmanabhan:1996ap,Tomboulis:1997gg,Modesto:2011kw,Biswas:2011ar,Modesto:2014lga,Buoninfante:2018mre,Buoninfante:2018gce} such as the Wataghin and Tomboulis-type form factors in \cref{sec:IntroNL} and the T-duality-inspired form factor in \cref{sec:IntroT}.

At leading logarithmic order, we obtain the bubble integral
\begin{equation}
 B\!\big(\tfrac{4\mu^2}{3}\big)
 =\frac{1}{8\pi^2}\left[
 \log\frac{\NLScale}{\max(m,\mu)}
 +\Ord(1)\right].
 \label{eq:BubbleLeadingLogThreshold}
\end{equation}
Physically, the bubble integral must depend on $\max(m,\mu)$ because the mass cuts off the logarithmic running in the IR such that the RG flow effectively freezes for $\mu<m$. Consequently, that regime cannot meaningfully contribute to the leading logarithm of $\lambda_{\rm pert}$ in \cref{eq:PerturbativeCriticalCoupling}. 

The weak triviality bound inherits the logarithm in \cref{eq:BubbleLeadingLogThreshold}. Plugging  \cref{eq:BubbleLeadingLogThreshold} into \cref{eq:PerturbativeCriticalCoupling} yields, at leading-logarithmic order,
\begin{align}
 \PhysicalCoupling_{\rm pert}(\mu;\PhysicalCoupling_{\rm crit})
 \approx&
 \frac{\PhysicalCoupling_{\rm crit}}{1
 +\frac{3\PhysicalCoupling_{\rm crit}}{16\pi^2}
 \log\frac{\NLScale}{\max(m,\mu)}},\\
\equiv&\PhysicalCoupling_{\rm pert,\log}(\mu;\PhysicalCoupling_{\rm crit})
.
 \label{eq:CriticalCouplingLeadingLog}
\end{align}
This bound exactly equals the EFT bound with perturbative ceiling $\PhysicalCoupling_{\rm crit}$ and cutoff $\NLScale$ \cite{Luscher:1987ay}. For fixed $m$, $\mu$, and $\PhysicalCoupling_{\rm crit}$, \cref{eq:CriticalCouplingLeadingLog} goes to zero as $\NLScale\to\infty$, recovering perturbative triviality. 

In summary, the leading logarithm of the weak triviality bound is independent of the specific form factor. We now turn to its universality.

\subsection{Scheme dependence}\label{sec:SchemeIndependence}

The coefficient of the leading logarithm is scheme-independent within the class of nonexceptional MOM configurations. For nonexceptional MOM configurations, ratios of the channel invariants $s$, $t$ and $u$ are bounded. This excludes the IR-singular exceptional points in the massless theory and prevents the subtraction prescription from approaching them in the low-energy limit. To parametrize a more general renormalization scheme, we introduce the parameters $\xi_s$, $\xi_t$ and $\xi_u$ such that
\begin{align}
	s=\xi_s\mu^2,&&t=\xi_t\mu^2,&&u=\xi_u\mu^2.
\end{align}
For example, $\xi_s=\xi_t=\xi_u=4/3$ recovers the symmetric-point subtraction we used in \cref{sec:MOM}. We further restrict to $p_I^2=\mu^2$, which implies that
\begin{align}
	\xi_s+\xi_t+\xi_u=4,\label{eq:SchemeParameters}
\end{align}
and require that $\xi_j>0$, \ie, that no channel has vanishing momentum. We repeat the derivation in \cref{sec:MOM,sec:WeakTriviality} through \cref{eq:PerturbativeCriticalCoupling} to obtain
\begin{equation}
	\PhysicalCoupling_{\rm pert}^{(\boldsymbol{\xi})}(\mu)=\frac{\PhysicalCoupling_{\rm crit}}{1+\frac{\PhysicalCoupling_{\rm crit}}{2}\sum_{j=s,t,u}B\left(\xi_j\mu^2\right)}.
\end{equation}
The leading logarithmic contribution is then
\begin{equation}
	\PhysicalCoupling_{\rm pert}^{(\boldsymbol{\xi})}(\mu)\approx\frac{\PhysicalCoupling_{\rm crit}}{1+\frac{\PhysicalCoupling_{\rm crit}}{16\pi^2}\sum_{j=s,t,u}\log\frac{\NLScale}{\max(m,\sqrt{\xi_j}\mu)}}.\label{eq:PertBoundScheme}
\end{equation}
Below the mass threshold, the mass sets the infrared scale in every channel. If $\mu\not\gg m$, the scheme parameters do not contribute significantly because $\sqrt{\xi_j}\mu\lesssim m$ for small $\xi_j$, \ie, for near-exceptional momentum configurations. Therefore, changing the fixed MOM configuration then affects only subleading terms.

Above the mass threshold, changing the MOM configuration adds a bounded term to the common logarithm. If $m\ll\mu$, we rewrite the logarithmic correction in \cref{eq:PertBoundScheme} as
\begin{equation}
	\begin{aligned}
		\sum_{j=s,t,u}\log&\frac{\NLScale}{\max(m,\sqrt{\xi_j}\mu)}\\
		&=3\log\frac{\NLScale}{\mu}+\sum_{j=s,t,u}\log\frac{1}{\max(m/\mu,\sqrt{\xi_j})}.\label{eq:PertBoundSchemeMassless}
	\end{aligned}
\end{equation}
Thus, the correction to the scheme-independent logarithm can only be large if one or two $\xi_j$ are very small. This means that the corresponding channels are soft. Quantitatively, the scheme parameters must satisfy
\begin{equation}
 \xi_s\xi_t\xi_u<(\mu/\NLScale)^6
 \label{eq:NearExceptionalCondition}
\end{equation}
for the renormalization scheme to contribute at the same order as the leading logarithm. Furthermore, the mass has to satisfy $m/\mu<(\mu/\NLScale)^{3/2}$ if two channels are soft and $m/\mu<(\mu/\NLScale)^{3}$ if one channel is soft. For example, for $\mu=\SI{246}{\giga\electronvolt}$, \ie, the electroweak scale, and $\NLScale=\MP$, $(\mu/\NLScale)^3\sim10^{-51}$. The corresponding bubble thus additionally probes the soft scale $\sqrt{\xi_j}\mu$, so $\mu$ alone no longer characterizes the momenta entering the subtraction prescription \cite{Poggio:1976qr,Ball:1994vf,Sturm:2009kb}. It is not specific to UV-finite theories that the leading logarithm is sensitive to near-exceptional configurations. Such a sensitivity also arises in the local theory, where large logarithms of ratios of Mandelstam variables affect the relation between MOM and $\overline{\mathrm{MS}}$ couplings and hence the inferred Landau-pole scale. 

In summary, the leading logarithm is universal across nonexceptional MOM configurations. It is also independent of the specific form factor. Therefore, we now study how much residual dependence on the form factor remains when evaluating the weak triviality bound exactly.

\subsection{Form-factor independence and phenomenology\label{sec:FF_dep}}

The weak triviality bound in \cref{eq:PerturbativeCriticalCoupling} depends on the specific form factor only at next-to-leading order. However, the leading order is only logarithmic and could in principle be overcome by form-factor-dependent finite corrections. We now show that the finite contributions from the Wataghin, Tomboulis-type, and T-duality-inspired form factors in \cref{sec:IntroNL,sec:IntroT}, as well as from a hard cutoff on the loop momentum, are small for the scale hierarchy relevant for present-day experiments and Planck-scale nonlocality. We expect that much smaller scale hierarchies yield less universal results because the finite contribution is only log-suppressed. 

\begin{figure}
	\includegraphics[width=\linewidth]{"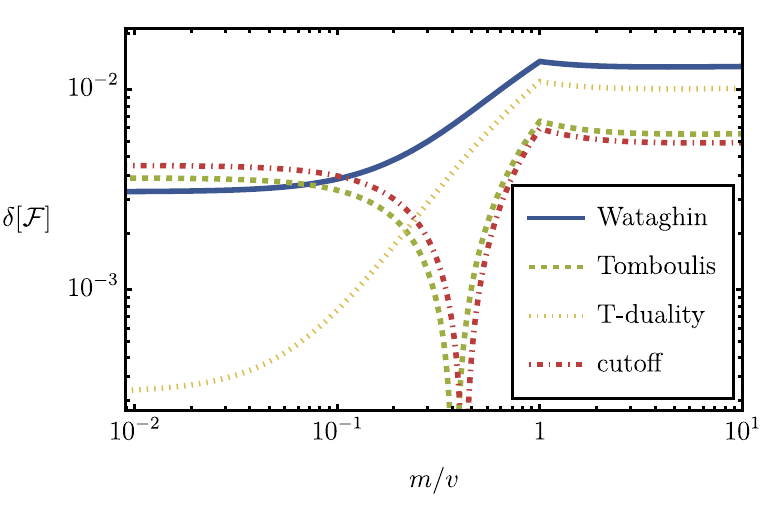"}
	\caption{We show the relative deviation $\delta[\FormFactor]$ of $\PhysicalCoupling_{\rm pert}$ from its leading-logarithmic approximation, as defined in \cref{eq:rel_dev}, as a function of the scalar mass for the three nonlocal form factors and a hard cutoff. We fix $\PhysicalCoupling_{\rm crit}=1$, $\mu=v=\SI{246}{\giga\electronvolt}$, and $\NLScale=\MP$.  \label{fig:relative_deviation}}
\end{figure}

We define the relative deviation from the leading logarithm for a form factor $\FormFactor$ as
\begin{equation}
	\delta[\FormFactor]\equiv\frac{|\PhysicalCoupling_{\rm pert}[\FormFactor]-\PhysicalCoupling_{\rm pert,\log}|}{\PhysicalCoupling_{\rm pert,\log}}.\label{eq:rel_dev}
\end{equation}
In \cref{fig:relative_deviation}, we plot the relative deviation for the Wataghin, Tomboulis-type and T-duality-inspired propagators introduced in \mbox{\cref{sec:IntroNL,sec:IntroT}} as well as a hard cutoff. We vary the mass $\Mass$ and fix $\PhysicalCoupling_{\rm crit}=1$ and $\mu=\SI{246}{\giga\electronvolt}$, and find that deviations are at most at the percent level. Thus, the choice of nonlocal propagator is largely irrelevant for the weak triviality bound, which, we recall, is essentially the local EFT triviality bound for a cutoff $\Lambda=\NLScale$. This bound is therefore a largely model-independent prediction of nonlocal QFTs, and likely also of models of discrete spacetime.

In the SM, triviality bounds constrain the Higgs mass. The Higgs mass is determined by the Higgs quartic coupling $\PhysicalCoupling_H$ and the electroweak scale, so the Higgs mass inherits the weak triviality bound. SM triviality bounds take into account coupled scalar, gauge, and Yukawa running and electroweak matching at two loops \cite{Hambye:1997ax,Ellis:2009tp}, yielding $M_H<\SI{175\pm1}{\giga\electronvolt}$ for the perturbative ceiling $\lambda_{H,{\rm crit}}=2\pi$ and a Planckian cutoff \cite{Ellis:2009tp}. Motivated by \cref{fig:relative_deviation}, we expect the form-factor-dependent corrections to this bound to be negligible. Thus, we expect it to apply to a wide class of nonlocal models, including models of discrete spacetime.

New physics may obstruct a perturbative nonlocal UV completion at the Planck scale. Additional particles and interactions change the beta functions and can lower the formal Landau-pole scale. If a coupling becomes strong well below $\NLScale$, the nonlocal suppression sets in too late to preserve perturbative control. The minimal supersymmetric $SO(10)$ model provides an example, with $\landau\sim\SI{1e17}{\giga\electronvolt}$ \cite{Aulakh:2003kg}. Consequently, a nonlocal supersymmetric $SO(10)$ model is likely ruled out because it cannot be UV complete.

\subsection{Scope and limitations\label{sec:ScopeLimitations}}

The weak triviality bound constrains a broad class of minimal-length models. We now discuss the limitations of our approach.

\textbf{Perturbation theory.} One cannot establish triviality in perturbation theory. We impose a finite UV coupling ceiling $\PhysicalCoupling_{\rm crit}$ to restrict the evolution to a chosen weak-coupling regime. Exceeding it does not prove triviality in the exact theory. However, without the nonlocal modifications, the exact theory is trivial \cite{Frohlich:1982tw,Aizenman:2019yuo,Luscher:1987ay}. If nonlocal effects remain negligible up to the Landau-pole scale $\landau$, we expect the Landau pole to persist.

\textbf{Off-shell result.} Our weak triviality bound is based on an off-shell renormalization scheme. The off-shell coupling is not invariant under field redefinitions. Consequently, imposing the same numerical ceiling in different field parametrizations need not impose the same physical condition. Similarly, beyond the leading logarithm in \cref{eq:CriticalCouplingLeadingLog}, the weak triviality bound depends on the renormalization scheme. A physical weak triviality bound constrains observables.

\textbf{Euclidean signature.} Our weak triviality bound is Euclidean. To constrain physical scattering observables, we need a consistent Lorentzian formulation. All three nonlocal propagators in \cref{sec:IntroNL,sec:IntroT} decay faster than allowed by a standard positive spectral representation with a particle pole. Ordinary Wick rotation may also fail because the nonlocal form factors can grow along complex loop-energy directions, obstructing the required contour deformation \cite{Buoninfante:2022krn}. Nonlocal theories with suitable entire form factors can nevertheless admit a perturbatively unitary Euclidean prescription followed by analytic continuation \cite{Efimov:1967pjn,Efimov:1969fd,Alebastrov:1972dxn,Pius:2016jsl,Chin:2018puw,Briscese:2018oyx,Briscese:2024tvc}. The leading logarithm in \cref{eq:CriticalCouplingLeadingLog} comes from loop momenta well below the nonlocality scale, where the propagators approach their local forms. Any consistent Lorentzian realization of the same low-energy scalar theory must therefore reproduce its coefficient.

\textbf{Other nonlocal models.} Our weak triviality bound neglects additional nonlocal interactions. We modify the propagator while keeping the quartic vertex local. Nonlocal modifications of the vertices which cannot be shifted to the propagator can change the UV behavior of the loop integrals and the resulting bound. We expect that the leading logarithm remains unchanged if those vertices deviate from the local $\phi^4$ interaction at the same nonlocality scale $\NLScale$.

\textbf{Form-factor constraints.} Our weak triviality bound does not apply to all possible nonlocal form factors. We assume that form factors are monotonic and sufficiently suppress UV degrees of freedom. The causal-set-inspired Euclidean propagator studied in \cite{deBrito:2023nie} is a counterexample. It approaches a nonzero constant at large momentum. This implies that the corresponding bubble in \cref{eq:bubble_integral} is not UV finite. Within the functional-RG approximation in \cite{deBrito:2023nie}, this modification accelerates the quartic running and brings the formal Landau pole close to the nonlocality scale. Thus, the effective nonlocal propagator triggers the Landau pole largely independently of the local Landau-pole scale.
 
\textbf{Extrapolation to SM.} In \cref{sec:FF_dep}, we assume that the weak dependence on the form factor found in the scalar theory extends to the full SM. Establishing this requires a consistent nonlocal prescription for fermions and gauge fields, including interactions compatible with gauge symmetry, and an explicit calculation of the resulting threshold corrections to the Higgs-mass bound. Nevertheless, we expect the local EFT approximation to remain accurate if the theory reproduces SM running well below $\NLScale$ and the transition to UV suppression occurs over a limited range of scales.

\section{Conclusion}\label{sec:concl}

UV divergences in loop integrals are not the only divergences that plague QFT. A central motivation for spacetime discreteness is to remove UV divergences, yet finite loop integrals do not automatically remove Landau poles. We show this at one loop in scalar $\phi^4$ theory with nonlocal form factors in the propagator. The form factors suppress the running beyond the fixed nonlocality scale $\NLScale$, but they prevent a breakdown of perturbation theory only if this suppression becomes effective before the Landau-pole scale $\landau$. At finite scales $\mu$, a finite range of interacting trajectories satisfies this constraint, which we call the weak triviality bound: interactions are allowed, but their strength is bounded. In the local limit, this range shrinks to zero and we recover ordinary triviality.

The weak triviality bound is largely independent of the underlying type of nonlocality. Its leading logarithm is universal: any construction that reproduces the same low-energy EFT inherits the same logarithmic running. The nonlocal form factor affects only the finite terms in the bound. For $\NLScale\sim\MP$, $\mu\sim v=\SI{246}{\giga\electronvolt}$, these terms change the bound at the percent level in the four examples we study, namely the Wataghin form factor, a Tomboulis-type and a T-duality-inspired form factor and a hard cutoff. We therefore expect the weak triviality bound to generalize to other nonlocal constructions such as models of discrete spacetime. 

Our result admits two different interpretations.
	
First, nonlocal UV completions may fail to resolve the UV problems they are intended to address. Nonlocal modifications can indeed render loop integrals finite, and gravitational models are perturbatively superrenormalizable or even finite \cite{Tomboulis:1997gg,Modesto:2014lga}. Yet, this only regularizes the divergences of perturbative Einstein gravity without renormalizing them. Landau poles are another divergence of perturbative QFT in the UV that nonlocality might not cure.

Second, the weak triviality bound is a useful constraint on the allowed theory space. Excluding trajectories that develop a pole restricts low-energy interactions and thereby increases the predictivity of nonlocal UV completions. For example, the EFT Higgs-mass bound obtained by requiring the SM Higgs quartic coupling to remain perturbative up to the Planck scale, \ie, $M_H<\SI{175\pm1}{\giga\electronvolt}$, becomes a constraint on nonlocal fundamental theories. However, this also highlights that the scale dependence of running couplings constrains nonlocal QFTs much less than local QFTs. A local QFT is only considered UV complete if it reaches a fixed point in the UV, which encodes infinitely many couplings in finitely many independent parameters. In nonlocal QFTs, weak triviality bounds restrict the allowed couplings, but there is no mechanism which could fix the infinite number of free parameters that the microscopic action can depend on.

Both interpretations weaken the case for nonlocality or discreteness as a fundamental principle for quantum gravity. The problem with the perturbative nonrenormalizability of quantum gravity is that it requires an infinite number of free parameters to be fixed in the action, not that it is UV divergent. A fundamental principle is required that reduces that number of parameters. Yet, discrete spacetime or nonlocal form factors cannot be that principle.

On a different note, one may seek a quantum-gravitational minimal length for reasons beyond removing UV divergences. Yet, a minimal-length scale may also emerge from a UV-complete continuum QFT. At a UV fixed point, the theory is scale-symmetric. If the physical scales that define measuring rods and clocks participate in this scaling, increasing the probing momentum may cease to improve spatial resolution relative to those scales. This could produce an operational minimal resolvable length scale around the crossover into the fixed-point regime \cite{Reuter:2005bb,Percacci:2010af}. Whether this mechanism is in fact realized, \eg, in asymptotically safe quantum gravity, remains unclear.

We see two challenges ahead. First, extending the calculation to a consistent nonlocal Standard Model would establish whether the finite regulator-dependent contributions remain negligible when gauge and Yukawa interactions are included. Second, our bound constrains a Euclidean off-shell coupling defined by momentum subtraction. UV suppression in Euclidean signature need not imply that Lorentzian scattering amplitudes are perturbative \cite{Koshelev:2021orf,Buoninfante:2022krn}. Therefore, it is crucial to obtain bounds on Lorentzian observables, such as scattering amplitudes.

\section*{Acknowledgments}

I thank Astrid Eichhorn for suggesting that I investigate a setup similar to \cite{deBrito:2023nie} in other models of quantum spacetime. Furthermore, I thank Benjamin Knorr for helpful discussions. I thank both Astrid Eichhorn and Benjamin Knorr for reading and commenting on the final version of the manuscript. I acknowledge the European Research Council’s (ERC) support under the European Union’s Horizon 2020 research and innovation program, grant agreement No. 101170215 (ProbeQG). This work was funded by the Federal Ministry of Education and Research (BMBF) and the Ministry of Science, Research and the Arts (MWK) Baden-Württemberg under the Excellence Strategy of the German federal and state governments. This work contributes to the COST Action CA23130 “Bridging high and low energies in search of quantum gravity (BridgeQG)”.

\appendix
\makeatletter
\renewcommand{\p@subsection}{\thesection.}
\makeatother

\section{Example form factors\label{app:examples}}

In \cref{sec:Setup}, we spell out the requirements we impose on form factors in this paper. Here, we introduce three example models that satisfy these requirements. The first two form factors are of Wataghin \cite{Wataghin:1934ann} and Tomboulis type \cite{Tomboulis:1997gg}, while the third follows from a T-duality-inspired zero-point length \cite{Padmanabhan:1996ap,Fontanini:2005ik}.

\subsection{Wataghin and Tomboulis-type form factors\label{sec:IntroNL}}

Non-degenerate higher-derivative kinetic terms generically introduce additional, potentially complex, poles in the propagator. These poles can imply Ostrogradski instabilities in the classical theory \cite{Ostrogradsky:1850,deUrries:1998obu,Woodard:2015zca} and obstruct unitarity when continuing Euclidean QFTs to Lorentzian signature \cite{Pais:1950za,Stelle:1976gc,deUrries:1998obu,Woodard:2015zca}.\footnote{Recent results show that classical Ostrogradski instabilities \cite{Deffayet:2023wdg,Held:2025fii} and unitarity violation in quantum theories \cite{Deffayet:2026uoe,Deffayet:2026cnu} can be avoided in specific models.} Wataghin and, especially, Tomboulis-type form factors are popular because they suppress large momenta without adding poles in the complex plane \cite{Tomboulis:1997gg,Modesto:2011kw,Biswas:2011ar,Modesto:2014lga,BasiBeneito:2022wux}.

Wataghin and Tomboulis-type form factors have the structure
\begin{equation}
	\FormFactor_p=e^{-H\left(p/\NLScale\right)},\label{eq:FormFactorNL}
\end{equation}
where $H$ is an entire function. The simplest example is the Wataghin form factor \cite{Wataghin:1934ann}
\begin{equation}
	H(p/\NLScale)=\Big(\frac{p}{\NLScale}\Big)^2.\label{eq:Wataghin}
\end{equation}
However, a form factor which suppresses UV modes exponentially may in fact not produce UV-finite gauge theories and gravity, because the nonlocal kinetic operators also generate momentum-dependent vertices \cite{Modesto:2017hzl}. Tomboulis-type form factors instead have polynomial UV asymptotics, which permit systematic power counting \cite{Tomboulis:1997gg,Modesto:2017hzl}; we choose the degree-two monomial example
\begin{equation}
	H(p/\NLScale)=\int_0^{(p/\NLScale)^4}\D z\frac{1-e^{-z^2}}{z},\label{eq:Tomboulis}
\end{equation}
also illustrated in \cite{BasiBeneito:2022wux}. The form factors characterized by \cref{eq:Wataghin} and \cref{eq:Tomboulis} have no poles in the complex plane and decrease monotonically. In the UV, the Tomboulis-type form factor characterized by \cref{eq:Tomboulis} goes like
\begin{equation}
	\FormFactor_p\sim\left(\frac{\NLScale}{p}\right)^4.\label{eq:TomboulisUV}
\end{equation}
In scalar $\phi^4$ theory, this falloff renders all loop integrals UV finite. We next introduce the third example: a T-duality-inspired propagator.

\subsection{T-duality-inspired form factor\label{sec:IntroT}}

T-duality motivates a zero-point length. For a string winding around a compact dimension of radius $R$, the spectrum is invariant under the transformation $R\leftrightarrow \alpha'/R$, where $\alpha'$ sets the string scale. Thus, distances below the string scale are mapped to distances above it, turning the string scale into an absolute minimum observable length. A qualitatively similar duality arises heuristically from the Compton wavelength, $M^{-1}$, and the Schwarzschild radius, $2l_{\rm P}^2M$. Their sum $M^{-1}+2l_{\rm P}^2M$ is invariant under the transformation $M^{-1}\leftrightarrow2l_{\rm P}^2M$, suggesting an analogous minimal length also from the interplay of gravity and quantum theory. We implement this idea by introducing a zero-point length in the scalar propagator.

We start with the worldline representation of the bare propagator \cite{Padmanabhan:1996ap}\footnote{In \cite{Padmanabhan:1996ap}, there is also a more rigorous derivation based on the discretized path integral for the free scalar.}
\begin{equation}
	\Prop_{\rm B}(x,x')=\int_0^\infty\frac{\D s}{(4\pi s)^2}\exp\left(-\frac{(x-x')^2}{4s}-\BareMass^2s\right).\label{eq:local_wordline_prop}
\end{equation}
We now replace $(x-x')^2\rightarrow(x-x')^2+\NLScale^{-2}$, \ie, we add a minimal-length scale $\NLScale^{-1}$ such that
\begin{equation}
	\Prop_{\rm B}(x,x')=\int_0^\infty\frac{\D s}{(4\pi s)^2}\exp\left(-\frac{(x-x')^2+\NLScale^{-2}}{4s}-\BareMass^2s\right).
\end{equation}
As shown in \cite{Padmanabhan:1996ap}, this interval shift is equivalent to modifying the QFT such that configurations related by the T-duality-like transformation $(x-x')^2\leftrightarrow\NLScale^{-4}/(x-x')^2$ are assigned equal weight in the path integral. In momentum space, the ensuing scalar propagator can be expressed as
\begin{equation}
	\Prop_{\rm B}(p)=\frac{1}{\NLScale\sqrt{p^2+\BareMass^2}}K_1\left(\NLScale^{-1}\sqrt{p^2+\BareMass^2}\right)\label{eq:PropTExpl},
\end{equation}
with the modified Bessel function of the second kind $K_n$. The corresponding form factor is
\begin{equation}
	\FormFactor_{{\rm B},p}=\NLScale^{-1}\sqrt{p^2+\BareMass^2}K_1\left(\NLScale^{-1}\sqrt{p^2+\BareMass^2}\right).\label{eq:FormFactorT}
\end{equation}
This form factor is monotonic for positive $p^2$, decays exponentially for large $p/\NLScale$ and has no poles in the complex plane. Unlike the form factors in \cref{sec:IntroNL}, however, it has a branch cut along $p^2\leq-\BareMass^2$.

\bibliographystyle{utphys} 
\bibliography{ref}

\end{document}